\documentclass[twocolumn,showpacs,amsmath,amssymb,superscriptaddress,pra]{revtex4}

\usepackage{graphicx}
\usepackage{dcolumn}
\usepackage{bm}
\usepackage{epsf}
\usepackage{amsmath}
\usepackage{amssymb}
\usepackage{color}

\newcommand{\be}{\begin{equation}}
\newcommand{\ee}{\end{equation}}
\newcommand{\bra}[1]{\langle #1|}
\newcommand{\ket}[1]{|#1\rangle}

\newcommand{\ketbrad}[1]{|#1\rangle\!\langle #1|}
\newcommand{\tr}[1]{\mathrm{tr}\left\{#1\right\}}

\newcommand{\la}{\left\langle}
\newcommand{\ra}{\right\rangle}

\newcommand{\de}[1]{\delta\left(#1\right)}
\newcommand{\td}{\mathrm{d}}

\newcommand{\e}[1]{\exp{\left(#1\right)}}

\newcommand{\id}{\mathbb{I}}

\newcommand{\bla}{bla\\bla\\bla\\bla\\bla}
\newcommand{\PR}{Phys. Rev.}
\newcommand{\PRA}{Phys. Rev. A}

\newcommand{\PRD}{Phys. Rev. D}
\newcommand{\PRE}{Phys. Rev. E}
\newcommand{\PRL}{Phys. Rev. Lett.}
\newcommand{\EPL}{EPL (Europhys. Lett.)}
\newcommand{\RMP}{Rev. Mod. Phys.}

\newcommand{\mc}[1]{\mathcal{#1}}
\newcommand{\mbb}[1]{\mathbb{#1}}

\newcommand{\mrm}[1]{\mathrm{#1}}

\begin{document}

\title{Holevo's bound from a general quantum fluctuation theorem}
\author{Dvir Kafri}
\affiliation{Joint Quantum Institute, Department of Physics, University of Maryland, College Park, MD 20742, USA}
\author{Sebastian Deffner}
\affiliation{Department of Chemistry and Biochemistry and Institute for Physical Sciences and Technology, University of Maryland, College Park, MD 20742, USA}

\date{\today}

\begin{abstract}
We give a novel derivation of Holevo's bound using an important result from nonequilibrium statistical physics, the fluctuation theorem. To do so we develop a general formalism of quantum fluctuation theorems for two-time measurements, which explicitly accounts for the back action of quantum measurements as well as possibly non-unitary time evolution. For a specific choice of observables this fluctuation theorem yields a measurement-dependent correction to the Holevo bound, leading to a tighter inequality. We conclude by analyzing equality conditions for the improved bound.
\end{abstract}

\pacs{05.30.-d, 05.40.-a, 89.70.Kn}

\maketitle

Thermodynamics and information theory are intimately connected. The most prominent evidence for this relationship is that the Clausius entropy \cite{clausius_64} is given by the Shannon information \cite{sha48} in systems at thermal equilibrium \cite{callen_85}. In particular, Landauer's principle \cite{lan61,ber12} illustrates that information is a physical, measurable quantity. Thermodynamic work has to be performed in order to create or erase information. Landauer's principle can therefore be understood as a statement of the second law of thermodynamics in an information theoretic context. This is also true for Holevo's bound \cite{hol98}, which limits the amount of classical information that can be encoded in a generic quantum system.  Recently, nano-devices - for which these principles are directly applicable - have become experimentally accessible \cite{mon02}. These controlled quantum systems have applications ranging from quantum simulation \cite{fey82,llo96}, cryptography \cite{eke91},  computing \cite{gro96,sho99,har09}, to metrology \cite{kas91,kim01,gio06,hud11}. The main obstacles towards realization of such devices are control noise and interactions with the environment. Thus, a thermodynamic study is necessary to fully understand their information theoretic properties. However, most nano-devices operate far from thermal equilibrium, so tools from non-equilibrium statistical physics are required. In recent years, formulations of the second law have been derived which are valid arbitrarily far from equilibrium. These so-called fluctuation theorems, in particular the Jarzynski equality \cite{jar97}, enable the calculation of equilibrium quantities from non-equilibrium averages over many realizations of a single process. They also encompass non-equilibrium, information theoretic generalizations of the second law. As Landauer's principle is a direct implication of this approach \cite{kaw07,jar08,esp11}, one may ask whether Holevo's bound is also such a result. One complication in this case is that the approach to fluctuation theorems for quantum systems is mathematically and conceptually more involved. Thermodynamic quantities, which are not given as state functions, cannot be assigned an Hermitian operator \cite{tal07}.  The proper formulation of quantum thermodynamics for non-equilibrium systems, especially quantum fluctuation theorems \cite{cam11}, must therefore be treated with care. 

The purpose of the present paper is twofold. In the first part we derive a general quantum fluctuation theorem that accounts for the back action of measurements on reduced systems (see \cite{ved12} for a similar approach). To this end, we consider an experimental point of view; we assume that the system of interest is coupled to an environment which is experimentally inaccessible. Such measurements on open quantum systems are inherently incomplete, since they ignore environmental degrees of freedom. Information is \textit{lost} that in principle could have been acquired by concurrent measurement of the reservoir. A general formulation of quantum fluctuation theorems must explicitly account for these effects. The integral fluctuation theorem we derive is applicable to arbitrary orthogonal measurements, for systems undergoing both unitary and non-unitary dynamics. 

In the second part we focus on an information theoretic consequence of the general quantum fluctuation theorem - Holevo's theorem. The derivation and implications of this result have attracted much attention \cite{yue93,hal93,fuc94,hau96,sch97}. Like more recent works \cite{sch962,jac03,jac06}, our derivation results in a sharpened statement of Holevo's bound, which takes into account the choice of measurement used to obtain the encoded information. Further our novel treatment is based on results weaker than the monotonicity of relative entropy, and directly leads to necessary and sufficient equality conditions.  This illustrates an interesting connection between quantum thermodynamics and quantum information theory.

\subparagraph{General quantum fluctuation theorem}

Consider a time-dependent quantum system, $\mc{S}$, with Hilbert space $\mc{H}_\mc{S}$ and initial density matrix $\rho_0$.  Information about the state of the system is obtained by performing measurements on $\mc{S}$ at the beginning and end of a specific process. Initially a quantum measurement is made of observable $A^\mrm{i}$, with eigenvalues $a_m^\mrm{i}$. Letting $\Pi_{m}^\mrm{i}$ denote the orthogonal projectors into the eigenspaces of $A^\mrm{i}$, we have $A^\mrm{i}=\sum_m a_m^\mrm{i}\Pi_m^\mrm{i}$. Note that the eigenvalues $\{a_m^\mrm{i}\}$ can be degenerate, so the projectors $\{\Pi_m^\mrm{i}\}$ may have rank greater than one. Unlike the classical case, as long as $\rho_0$ and $A^\mrm{i}$ do not have a common set of eigenvectors - i.e. they do not commute - performing a measurement on $\mc{S}$ alters its statistics. Measuring $a_m^\mrm{i}$ maps $\rho_0$ to the state $\Pi_m^\mrm{i} \rho_0\Pi_m^\mrm{i}/p_m$, where $p_m = \tr{ \Pi_m^\mrm{i} \rho_0\Pi_m^\mrm{i}}$ is the probability of the measurement outcome $a_m^\mrm{i}$. Generally accounting for all possible measurement outcomes, the statistics of $\mc{S}$ after the measurement are given by the weighted average of all projections, 
\begin{equation}
\label{e02}
M^\mrm{i}(\rho_0) = \sum_m \Pi_m^\mrm{i}\, \rho_0\, \Pi_m^\mrm{i}\,.
\end{equation} If $\rho_0$ commutes with $A^\mrm{i}$, it commutes with each $\Pi_m^\mrm{i}$, so $M^\mrm{i}(\rho_0) =\sum_m \Pi_m^\mrm{i} \Pi_m^\mrm{i} \rho_0  =\rho_0$ and the statistics of the system are unaltered by the measurement. After measuring $a_m^\mrm{i}$, $\mc{S}$ undergoes a generic time evolution, after which it is given by $\mbb{E}( \Pi_m^\mrm{i} \rho_0 \Pi_m^\mrm{i})/p_m$. Here $\mbb{E}$ represents any linear (unitary or non-unitary) quantum transformation, which is trace-preserving and maps non-negative operators  to non-negative operators. Further, we require that this holds whenever $\mbb{E}$ is extended to an operation $\mbb{E} \otimes \id_\mc{E}$ on any  enlarged Hilbert space $\mc{H}_S\otimes\mc{H}_\mc{E}$ ($\id_\mc{E}$ being the identity map on $\mc{H}_\mc{E}$). Such a transformation is called a trace-preserving, completely positive (TCP) map \cite{sch96}. After this evolution, a measurement of a second (not necessarily the same) observable, $A^\mrm{f} = \sum_n a_n^\mrm{f} \Pi_{n}^\mrm{f}$, is performed on $\mc{S}$. The probability of measuring $a_n^\mrm{f}$, conditioned on having first measured $a_m^\mrm{i}$, is $p_{n|m} =\tr{\Pi_n^\mrm{f}\, \mbb{E} \left( \Pi_m^\mrm{i} \rho_0\Pi_m^\mrm{i}\right)}/p_m$. Accordingly, the joint probability distribution $p_{m\rightarrow n}$ reads
\begin{equation}
\label{e03}
p_{m\rightarrow n} = p_m \cdot p_{n|m} = \tr{\Pi_n^\mrm{f}\, \mbb{E} \left( \Pi_m^\mrm{i} \rho_0\Pi_m^\mrm{i}\right)}\,.
\end{equation}
We are interested in the probability distribution of possible measurement outcomes, $\mathcal{P}\left( \Delta a\right)=\la \de{\Delta a-\Delta a_{n,m}}\ra$, where $\Delta a_{n,m} = a_n^\mrm{f} - a_m^\mrm{i}$ is a random variable determined in a single measurement run. Its probability distribution is given by averaging over all possible realizations,
\begin{equation}
\label{e04}
\mathcal{P}\left( \Delta a\right)=\sum\limits_{m,n} \de{\Delta a-\Delta a_{n,m}}\, p_{m\rightarrow n}\,.
\end{equation}
To derive the integral fluctuation theorem we follow the standard approach and compute its characteristic function, $\mc{G}(s)$, which is the Fourier transform of $\mc{P}(\Delta a)$ \cite{cam11}
\begin{equation}
\label{e05}
\begin{split}
\mc{G}(s)& = \int \td(\Delta a)\,\mc{P}(\Delta a)\,\e{i s\,\Delta a} \\
& = \tr{ \e{i s A^\mrm{f}}\,  \mbb{E} \left( M^\mrm{i}(\rho_0) \e{-i s A^\mrm{i}}\right)  }\,.
\end{split}
\end{equation}
Choosing $s = i$, we obtain the general quantum fluctuation theorem 
\begin{equation}
\label{e06}
\la \e{- \Delta a} \ra = \gamma\,.
\end{equation}
Since it is explicitly dependent on the map $\mbb{E}$, the quantity $\gamma$ accounts for the  information \textit{lost} by not measuring the environment. It plays a crucial role in the following discussion and is given by
\begin{equation}
\label{e07}
\gamma = \tr{\e{-A^\mrm{f}}\, \mbb{E} \left(M^\mrm{i}(\rho_0) \e{ A^\mrm{i}} \right) }\,.
\end{equation}
Similar fluctuation theorems of the form $\la \e{-\Sigma}\ra=\gamma_\mrm{cl}$ have been derived in the context of classical feedback processes, where $\Sigma$ is an entropy production \footnote[1]{For these the right hand side, $\gamma_\mrm{cl}$, is commonly called the (classical) efficacy of the feedback protocol \cite{sag10,mor11}. In the present context $\gamma$ can  then be interpreted as  a quantum efficacy corresponding to the observables $A^\mrm{i}$ and $A^\mrm{f}$, and the TCP map $\mbb{E}$. Note that $\gamma$ only takes the from of the classical efficacy, and usually may not be regarded as its physical quantum analogue \cite{sag12}.}. We note that by appropriate choice of initial and final observables  $A^\mrm{i}$ and $A^\mrm{f}$, Eq.~\eqref{e07} reproduces many known quantum fluctuation theorems \cite{tal07,tal09,ved12,cam11,def11,cam11a,woj04,mor11}, which we will discuss in detail elsewhere.

 A complementary result to the fluctuation theorem is Jensen's Inequality, which states that for any convex function $\phi''(x)\geq 0$ and random variable $x$, $\la\phi(x)\ra \geq \phi(\la x\ra)$ \cite{chandler_87}. Applying this to Eq.~\eqref{e06} yields 
\begin{equation}
\label{e08}
 \la \Delta a\ra \geq -\ln{(\gamma)}\,. 
\end{equation}
For specific choices of thermodynamically relevant observables $A^\mrm{i}$ and $A^\mrm{f}$, this relation can be understood as a formulation of the Clausius inequality. In particular, for a unitary time-evolution $U_\tau = \mc{T}_> \e{-i \int_0^\tau H(t) \mbox{d}t}$, an initial Gibbsian state $\rho_0=\e{-\beta H(0)}/Z_0$ and corresponding energy measurements, $A^\mrm{i}=\beta H(0)$ and $A^\mrm{f}=\beta H(\tau)$, Eq.~\eqref{e06} re-produces the quantum Jarzynski equality \cite{kur00,tas00,tal07}. Accordingly, Eq.~\eqref{e08} reduces  to the maximum work theorem, $\beta \la W\ra\geq \beta \Delta F$, where $\la W\ra=\la H(\tau) \ra-\la H(0)\ra = \la \Delta a \ra/\beta$ and $\beta \Delta F= -\ln{\left(Z_\tau/Z_0\right)} = - \ln(\gamma)$. 

\subparagraph{Holevo's bound}
We now use the fluctuation theorem \eqref{e06} to derive a sharpened version of Holevo's bound. This bound sets a limit on how much classical information can be sent through a (noisy) quantum channel. Let us consider a message composed of code words $w_j$ that appear with probability $\pi_j$.  A messenger (Alice) attempts to transfer this message to a receiver (Bob) by encoding each word $w_j$ in a quantum state and transmitting that state to Bob. We assume that Bob receives the state $\rho_j$, which may have come through a lossy medium and therefore be different from the original state prepared by Alice. Bob attempts to infer the word $w_j$ from the encoding by making a generalized measurement of the state $\rho_j$. This corresponds to introducing a probe, initially in a pure state $\ket{0}$, and making an orthogonal measurement on the compound state $\rho_j \otimes \ket{0}\bra{0}$ \cite{bra92}. If $\{\Pi_k\}$ represents the set of orthogonal projectors corresponding to Bob's measurement, the probability of measuring $\Pi_k$, given message $w_j$, is given by
\begin{equation}
\label{e09}
\pi_{k | j} = \tr{\rho_j \otimes \ket{0}\bra{0} \Pi_k}= \tr{\rho_j M_k}\,,
\end{equation}
where $M_k = \bra{0} \Pi_k \ket{0}$ are operators acting only on the encoding degree of freedom. Although the operators $\{M_{k}\}$ are non-negative and $\sum_k M_k = \id$, they are generally  not projectors, $M_k^2 \neq M_k$. Such a collection $\{M_k\}$ is called a positive operator valued measure (POVM), and describes the most general measurement on a quantum system. The classical message distribution $\{\pi_j\}$, output quantum encoding $\{\rho_j\}$, and POVM elements $\{M_k\}$ define a classical-quantum channel \cite{nielsen_00}.

 A proper measure of how well Bob decodes Alice's message is the mutual information between the encoded message and measurement distributions, $I = \sum_{j k} \pi_j \, \pi_{k|j}\, \ln{\left(\pi_{k|j}/\pi_k \right)}$, where $\pi_k = \sum_j \pi_j\, \pi_{k|j}$ is the overall probability of  measuring $\Pi_k$. Note that $I = \sum_j \pi_j D(\pi_{k|j}||\pi_k)$, where $D(\pi_{k|j}||\pi_k) =\sum_k \pi_{k|j} \ln\left(\pi_{k|j}/\pi_k \right) $ is the (classical) relative entropy \cite{kullback_78}.  Hence $I$ is a sum of non-negative terms and is $0$ if and only if $\pi_{k|j} = \pi_k $ for all $k,j$. That is, $I$ vanishes only if all outcomes of the measurement are independent of the encoded word, so that Bob always learns nothing about the message. 
 
 The probability of the message being $w_j$, conditioned on Bob measuring $\Pi_k$ is, $\pi_{j | k} = \pi_{k|j}\, \pi_j /\pi_{k}$. We have
\begin{equation}
\label{e10}
I = S(\{\pi_j\}) + \sum_k \pi_k \sum_j \pi_{j|k} \ln{\pi_{j|k}},
\end{equation}
where $S(\{\pi_j\}) = -\sum_j \pi_j \ln \pi_j$ is the Shannon information of distribution $\{\pi_j\}$.  Since $x \ln(x) \leq 0$ for $|x|\leq 1$, with equality only for $x = 0$ and $x=1$, we observe that $I\leq S(\{\pi_j\})$, with equality if and only if $\pi_{j|k}$ is $0$ or $1$ for all $j,k$. In other words, the mutual information $I$ is at most $S(\{\pi_j\})$, with equality if and only if Bob correctly decodes the message in every instance. If Alice's encoded states are not perfectly distinguishable (that is, if the supports of $\rho_j$ and $\rho_{j'}$ are not orthogonal for some $j\neq j'$), then $I$ can never equal $S(\{\pi_j\})$, no matter what measurement Bob chooses to make.  Holevo's theorem is then an upper bound for $I$, namely
\begin{equation}
\label{e11}
\chi \equiv S(\bar \rho) - \sum_j \pi_j \cdot S(\rho_j) \geq I,
\end{equation}
where $S(\rho) = -\tr{\rho \ln \rho}$ is the von-Neumann entropy, and $\bar \rho = \sum_{j} \pi_j \rho_j$ is the density matrix describing the statistics of the encoding given no knowledge of the message word. Heuristically,  the Holevo quantity $\chi$ can be considered as the uncertainty of the encoding with no knowledge of the message, minus the average remaining uncertainty given knowledge of the message. Note that if the encoded states are distinguishable, i.e. $\rho_j \rho_{j'} = \mbb{O}$ for all $j \neq j'$, then $\chi = S(\{\pi_j\})$, so that with a proper measurement Bob may always correctly decode Alice's message.

We now show how Holevo's theorem \eqref{e11} follows  as a consequence of the general quantum fluctuation theorem \eqref{e06}. To do this we must appropriately choose initial state, evolution operation, and observables so that the random variable $\la\Delta a\ra$ averages to $\chi - I$. In the language of the general fluctuation theorem, let the initial state $\rho_0$ reside in a composite Hilbert space $\mc{H}_\mc{E}\otimes \mc{H}_\mc{P}\otimes \mc{H}_\mc{M}$. $\mc{H}_\mc{E}$ represents the encoding Hilbert space, which Alice prepares and Bob then measures, $\mc{H}_\mc{P}$ is the probe Hilbert space accessible only to Bob, and $\mc{H}_\mc{M}$ is a \textit{message} Hilbert space. Note that $\mc{H}_\mc{M}$ is not a \textit{real}, physically accessible subspace, but rather a mathematical construction denoting the \textit{memory} for the classical information of the message \cite{nielsen_00}. We have
\begin{equation}
\label{e12}
\rho_0 = \sum_j \pi_j \rho_j \otimes \ket{0}\bra{0}\otimes  \ket{j}\bra{j}\,,
\end{equation}
where the states $\ket{j}$, each corresponding to word $w_j$, form an orthonormal basis for $\mc{H}_\mc{M}$. We see that, with probability $\pi_j$, $\rho_0$ corresponds to the \textit{message state} $\ket{j}$. The measured initial and final observables are
\begin{equation}
\label{e13}
\begin{split}
A^\mrm{i} &= \sum_j \ln\left(   \hat \rho_j^{-1} \right)  \otimes \ket{0}\bra{0}\otimes \ket{j}\bra{j}\\
A^\mrm{f} &= - \ln\left(\bar \rho \otimes \ket{0}\bra{0}\right) \otimes \id_\mc{M}  \\
& \quad -  \sum_{k,j}I_{k,j}  \Pi_k\otimes \ket{j}\bra{j}\,,
\end{split}
\end{equation}
 where $I_{k,j}=\ln{\left(\pi_{k|j}/\pi_{k}\right)}$. Here $\hat \rho_j^{-1}$ denotes the inverse within the support of $\rho_j$, so that $\hat \rho_j^{-1} \ket{\psi} = 0$ whenever $\rho_j \ket{\psi} = 0$. This form for Eq.~\eqref{e13} ensures that $\exp(A^\mrm{i})$ and $\exp(-A^\mrm{f})$ are bounded operators. 

Note that the states $\rho_j$ do not represent the original encoding set up by Alice, but rather its time evolved state after undergoing dynamics in a quantum channel. To apply the fluctuation theorem \eqref{e06}, we start with the output of this channel and perform the two measurements, $A^\mrm{i}$ and $A^\mrm{f}$, immediately after each other. The TCP map crucial for Eq.~\eqref{e06} is thus the identity map $\mbb{E}(\rho)=\rho$. $A^\mrm{i}$ commutes with $\rho_0$, so after measurement of $A^\mrm{i}$, measurement of $A^\mrm{f}$ is carried out on the \textit{same} state, $M^\mrm{i}(\rho_0) = \rho_0$. Computing $ \la\Delta a\ra = \tr{(A^\mrm{f}-A^\mrm{i})\rho_0} = \chi - I$, Eq.~\eqref{e08} is
\begin{equation}
\label{e14}
\chi - I \geq -\ln{\left(\gamma\right)}\,,
\end{equation}
where the corresponding quantum efficacy  is given by
\begin{equation}
\label{e15}
\gamma = \tr{\e{-A^\mrm{f}} \rho_0 \e{A^\mrm{i}} }\,.
\end{equation}
Equations \eqref{e14} and \eqref{e15} constitute the sharpened Holevo's bound as a consequence of the general quantum fluctuation theorem \eqref{e06}. Indeed, our new bound is tighter than the usual inequality \eqref{e11}, in the sense that the correction term, $-\ln{(\gamma)}$, is always non-negative. Consider 
\begin{equation}
\label{e16}
\begin{split}
\rho_0\, \e{A^\mrm{i}} &= \sum_j \pi_j \rho_j \e{\ln(\hat \rho_j^{-1})} \otimes \ket{0}\bra{0}\otimes \ket{j}\bra{j}\\
& =  \sum_j \pi_j \hat P_j \otimes \ket{0}\bra{0}\otimes \ket{j}\bra{j} \,,
\end{split}
\end{equation}
where $\hat P_j$ is the projector into the support of  $\rho_j$.  We can rewrite Eq.~\eqref{e15} with Eq.~\eqref{e16} as
\begin{equation}
\label{e17}
\begin{split}
&\gamma = \tr{\e{-A^\mrm{f}}\, \rho_0\, \e{A^\mrm{i}}}\\
&= \mrm{tr}\Bigg\{ \e{-A^\mrm{f}} \sum_j \pi_j \hat P_j  \otimes \ket{0}\bra{0} \otimes \ket{j}\bra{j} \Bigg\}\\
&  \leq  \sum_j \pi_j\, \tr{ \e{\ln\left(\bar \rho\otimes \ket{0}\bra{0}\right) + \sum_{k}I_{k,j}\,  \Pi_k  }}\,.
\end{split}
\end{equation}
where the inequality is justified by noting that $\e{\ln\left(\bar \rho\otimes \ket{0}\bra{0}\right) + \sum_{k}I_{k,j}\, \Pi_k  }$ is non-negative and $ \hat P_j \otimes \ket{0}\bra{0}$ is a projection operator. We now use a statement of the Golden-Thompson inequality \cite{gol65,tho65}, that is for any Hermitian operators $A$ and $B$, we have $\tr{\e{A+B} } \leq \tr{\e{A}\e{B} }$. Note that in the present case, $A$ and $B$ are both logarithms of bounded Hermitian operators, and are only bounded from \textit{above}, though the Golden-Thompson inequality still holds \cite{rus72}. Accordingly, we have
\begin{equation}
\label{e18}
\begin{split}
&\gamma=\tr{\e{-A^\mrm{f}}\, \rho_0\, \e{A^\mrm{i}}}\\
& \leq  \sum_j \pi_j\, \tr{ \e{\ln\left(\bar \rho\otimes \ket{0}\bra{0}\right) }\cdot \e{ \sum_{k}I_{k,j} \, \Pi_k } }\\
& =  \sum_j \pi_j\, \tr{ \left(\bar \rho\otimes \ket{0}\bra{0}\right) \sum_{k}\pi_{k|j}/\pi_k\,  \Pi_k  }\,.
\end{split}
\end{equation}
From the definition $\pi_k=\sum_j \pi_j\pi_{k|j}$ we finally obtain
\begin{equation}
\label{e18a}
\gamma\leq    \tr{\bar \rho \otimes \ket{0}\bra{0}  \sum_k \Pi_k}= 1\,,
\end{equation}
which shows that $-\ln{(\gamma)} \geq 0$, as desired. We note that our derivation does not invoke the monotonicity of the relative entropy or equivalent statements \cite{rus02}. Instead we have used only Jensen's inequality and the Golden-Thompson inequality, which are weaker results \cite{pet88,pet94}.

\paragraph*{Equality conditions}
Holevo's bound \eqref{e13} is obtained with the help of Jensen's inequality. For strictly convex functions $\phi''(x)>0$, the Jensen bound $\la \phi(x)\ra \geq \phi(\la x\ra )$ achieves equality if and only if the random variable $x$ is constant valued. This allows us to derive the equality conditions for \eqref{e14} in a straightforward manner. Specifically, equality is achieved  only if  
\begin{equation}
\label{e19}
 \left(-\ln(\hat \rho_j^{-1}) - \ln(\bar \rho) - \sum_k I_{k,j} M_k \right)\hat P_j
 = - \ln(\gamma) \hat P_j
\end{equation}
for all $j$. This follows from a few simple observations.

First assume that $\chi-I = -\ln(\gamma)$.  As $A^\mrm{i}$, $\rho_0$ and the projectors $\id\otimes\ketbrad{0}\otimes\ketbrad{j}$ mutually commute, we consider a mutual eigenprojector $R_{m j} = R_m^{(j)} \otimes \ketbrad{0}\otimes \ketbrad{j}$ such that $A^\mrm{i}R_{m j}= a_m^\mrm{i} R_{mj}$ and $\rho_0 R_{mj} \neq 0$.  Since the function $\e{x}$ is strictly convex, the random variable $\Delta a_{n m}$ obtained from the measurements of $A^\mrm{i}$ and $A^\mrm{f}$ has to satisfy $\Delta a_{n m}= -\ln(\gamma)$ for all measurements with nonzero probability. Hence an initial measurement of $a_m^\mrm{i}$ implies with certainty a final measurement $a_m^\mrm{i} - \ln(\gamma)$. Since $R_{m j}$ is a projector into an eigenspace of $\rho_0$, any state satisfying $R_{m j} \ket{\psi} = \ket{\psi}$ must therefore also be an eigenstate of $A^\mrm{f}$ with eigenvalue $a_m^\mrm{i} - \ln(\gamma)$, so
\begin{equation}
\label{e20}
(A^\mrm{f} -A^\mrm{i})R_{mj} =  - \ln(\gamma)R_{mj}.
\end{equation}
Using the definition \eqref{e13} and $M_k = \bra{0}\Pi_k \ket{0}$, Eq.~\eqref{e19} follows by summing on $m$ noting that $\sum_m R_{m j}= \hat P_j\otimes \ketbrad{0}\otimes \ketbrad{j}$.

Conversely, assume that Eq.~\eqref{e19} holds for all $j$. Since $A^\mrm{i}$ and $\rho_0$ commute, we have
\begin{equation}
\label{e21}
\begin{split}
&\chi - I  =  \tr{\rho_0 (A^\mrm{f} - A^\mrm{i})}\\
& = \sum_j \pi_j \tr{\rho_j \left( \ln(\rho_j)-\ln(\bar \rho) - \sum_k I_{k,j} M_k \right) \hat P_j} \\
& = \sum_j \pi_j \tr{\rho_j \left(-\ln(\gamma) \hat P_j\right)} = -\ln(\gamma)\sum_j \pi_j \\
& =  -\ln(\gamma)
\end{split}
\end{equation}
We conclude that Eq.~\eqref{e19} is equivalent to equality in Eq.~\eqref{e14}.  Observe that since $\chi - I \geq -\ln(\gamma) \geq 0$, the equality condition for $\chi = I$,  Eq.~\eqref{e19} with $\ln(\gamma)=0$, is obtained as a corollary of our result \cite{rus02}. The equality condition \eqref{e19} may be used to determine the bound saturating observable $A^\mrm{f}$ self-consistently.

\subparagraph{Concluding remarks}
We developed a general framework for quantum fluctuation theorems by explicitly accounting for the back action of quantum measurements. With this new result, we showed that quantum mechanical formulations of the second law are intimately tied to quantum information theory by deriving Holevo's bound as a consequence of a fluctuation theorem. The new approach not only provides a simple derivation, but also a sharpened statement of the original bound and a corresponding equality criterion. 

\acknowledgments{The authors thank Jacob Taylor and Eric Lutz for interesting discussions. SD acknowledges financial support by a fellowship within the postdoc-program of the German Academic Exchange Service (DAAD, contract No D/11/40955). DK acknowledges financial support by a fellowship from the Joint Quantum Institute.}


\begin{thebibliography}{52}
\expandafter\ifx\csname natexlab\endcsname\relax\def\natexlab#1{#1}\fi
\expandafter\ifx\csname bibnamefont\endcsname\relax
  \def\bibnamefont#1{#1}\fi
\expandafter\ifx\csname bibfnamefont\endcsname\relax
  \def\bibfnamefont#1{#1}\fi
\expandafter\ifx\csname citenamefont\endcsname\relax
  \def\citenamefont#1{#1}\fi
\expandafter\ifx\csname url\endcsname\relax
  \def\url#1{\texttt{#1}}\fi
\expandafter\ifx\csname urlprefix\endcsname\relax\def\urlprefix{URL }\fi
\providecommand{\bibinfo}[2]{#2}
\providecommand{\eprint}[2][]{\url{#2}}

\bibitem[{\citenamefont{Clausius}(1864)}]{clausius_64}
\bibinfo{author}{\bibfnamefont{R.}~\bibnamefont{Clausius}},
  \emph{\bibinfo{title}{Abhandlungen \"uber die mechanische {W}\"armetheorie}}
  (\bibinfo{publisher}{Vieweg}, \bibinfo{address}{Braunschweig, Germany},
  \bibinfo{year}{1864}).

\bibitem[{\citenamefont{Shannon}(1948)}]{sha48}
\bibinfo{author}{\bibfnamefont{C.~E.} \bibnamefont{Shannon}},
  \bibinfo{journal}{Bell Sys. Tech. J.} \textbf{\bibinfo{volume}{27}},
  \bibinfo{pages}{379} (\bibinfo{year}{1948}).

\bibitem[{\citenamefont{Callen}(1985)}]{callen_85}
\bibinfo{author}{\bibfnamefont{H.}~\bibnamefont{Callen}},
  \emph{\bibinfo{title}{Thermodynamics and an Introduction to
  Thermostastistics}} (\bibinfo{publisher}{Wiley}, \bibinfo{address}{New York,
  USA}, \bibinfo{year}{1985}).

\bibitem[{\citenamefont{Landauer}(1961)}]{lan61}
\bibinfo{author}{\bibfnamefont{R.}~\bibnamefont{Landauer}},
  \bibinfo{journal}{IBM J. Research and Develop.} \textbf{\bibinfo{volume}{5}},
  \bibinfo{pages}{183} (\bibinfo{year}{1961}).

\bibitem[{\citenamefont{B\'{e}rut et~al.}(2012)\citenamefont{B\'{e}rut,
  Arakelyan, Petrosyan, Ciliberto, Dillenscheinder, and Lutz}}]{ber12}
\bibinfo{author}{\bibfnamefont{A.}~\bibnamefont{B\'{e}rut}},
  \bibinfo{author}{\bibfnamefont{A.}~\bibnamefont{Arakelyan}},
  \bibinfo{author}{\bibfnamefont{A.}~\bibnamefont{Petrosyan}},
  \bibinfo{author}{\bibfnamefont{S.}~\bibnamefont{Ciliberto}},
  \bibinfo{author}{\bibfnamefont{R.}~\bibnamefont{Dillenscheinder}},
  \bibnamefont{and} \bibinfo{author}{\bibfnamefont{E.}~\bibnamefont{Lutz}},
  \bibinfo{journal}{Nature} \textbf{\bibinfo{volume}{483}},
  \bibinfo{pages}{187} (\bibinfo{year}{2012}).

\bibitem[{\citenamefont{Holevo}(1998)}]{hol98}
\bibinfo{author}{\bibfnamefont{A.~S.} \bibnamefont{Holevo}},
  \bibinfo{journal}{IEEE Trans. Info. Theo.} \textbf{\bibinfo{volume}{44}},
  \bibinfo{pages}{269} (\bibinfo{year}{1998}).

\bibitem[{\citenamefont{Monroe}(2002)}]{mon02}
\bibinfo{author}{\bibfnamefont{C.}~\bibnamefont{Monroe}},
  \bibinfo{journal}{Nature} \textbf{\bibinfo{volume}{416}},
  \bibinfo{pages}{238} (\bibinfo{year}{2002}).

\bibitem[{\citenamefont{Feynman}(1982)}]{fey82}
\bibinfo{author}{\bibfnamefont{R.}~\bibnamefont{Feynman}},
  \bibinfo{journal}{Int. J. Theo. Phys.} \textbf{\bibinfo{volume}{21}},
  \bibinfo{pages}{476} (\bibinfo{year}{1982}).

\bibitem[{\citenamefont{Lloyd}(1996)}]{llo96}
\bibinfo{author}{\bibfnamefont{S.}~\bibnamefont{Lloyd}},
  \bibinfo{journal}{Science} \textbf{\bibinfo{volume}{273}},
  \bibinfo{pages}{1073} (\bibinfo{year}{1996}).

\bibitem[{\citenamefont{Ekert}(1991)}]{eke91}
\bibinfo{author}{\bibfnamefont{A.~K.} \bibnamefont{Ekert}},
  \bibinfo{journal}{\PRL} \textbf{\bibinfo{volume}{67}}, \bibinfo{pages}{661}
  (\bibinfo{year}{1991}).

\bibitem[{\citenamefont{Grover}(1996)}]{gro96}
\bibinfo{author}{\bibfnamefont{L.~K.} \bibnamefont{Grover}},
  \bibinfo{journal}{Annual ACM Symposium of Theory of Computing} p.
  \bibinfo{pages}{212} (\bibinfo{year}{1996}).

\bibitem[{\citenamefont{Shor}(1999)}]{sho99}
\bibinfo{author}{\bibfnamefont{P.~W.} \bibnamefont{Shor}},
  \bibinfo{journal}{SIAM Review} \textbf{\bibinfo{volume}{41}},
  \bibinfo{pages}{303} (\bibinfo{year}{1999}).

\bibitem[{\citenamefont{Harrow et~al.}(2009)\citenamefont{Harrow, Hassidim, and
  Lloyd}}]{har09}
\bibinfo{author}{\bibfnamefont{A.~W.} \bibnamefont{Harrow}},
  \bibinfo{author}{\bibfnamefont{A.}~\bibnamefont{Hassidim}}, \bibnamefont{and}
  \bibinfo{author}{\bibfnamefont{S.}~\bibnamefont{Lloyd}},
  \bibinfo{journal}{\PRL} \textbf{\bibinfo{volume}{103}},
  \bibinfo{pages}{150502} (\bibinfo{year}{2009}).

\bibitem[{\citenamefont{Kasevich and Chu}(1991)}]{kas91}
\bibinfo{author}{\bibfnamefont{M.}~\bibnamefont{Kasevich}} \bibnamefont{and}
  \bibinfo{author}{\bibfnamefont{S.}~\bibnamefont{Chu}},
  \bibinfo{journal}{\PRL} \textbf{\bibinfo{volume}{67}}, \bibinfo{pages}{181}
  (\bibinfo{year}{1991}).

\bibitem[{\citenamefont{Kimble et~al.}(2001)\citenamefont{Kimble, Levin,
  Matsko, Thorne, and Vyatchanin}}]{kim01}
\bibinfo{author}{\bibfnamefont{H.~J.} \bibnamefont{Kimble}},
  \bibinfo{author}{\bibfnamefont{Y.}~\bibnamefont{Levin}},
  \bibinfo{author}{\bibfnamefont{A.~B.} \bibnamefont{Matsko}},
  \bibinfo{author}{\bibfnamefont{K.~S.} \bibnamefont{Thorne}},
  \bibnamefont{and} \bibinfo{author}{\bibfnamefont{S.~P.}
  \bibnamefont{Vyatchanin}}, \bibinfo{journal}{\PRD}
  \textbf{\bibinfo{volume}{65}}, \bibinfo{pages}{022002}
  (\bibinfo{year}{2001}).

\bibitem[{\citenamefont{Giovannetti et~al.}(2006)\citenamefont{Giovannetti,
  Lloyd, and Maccone}}]{gio06}
\bibinfo{author}{\bibfnamefont{V.}~\bibnamefont{Giovannetti}},
  \bibinfo{author}{\bibfnamefont{S.}~\bibnamefont{Lloyd}}, \bibnamefont{and}
  \bibinfo{author}{\bibfnamefont{L.}~\bibnamefont{Maccone}},
  \bibinfo{journal}{\PRL} \textbf{\bibinfo{volume}{96}},
  \bibinfo{pages}{010401} (\bibinfo{year}{2006}).

\bibitem[{\citenamefont{Hudson et~al.}(2011)\citenamefont{Hudson, Kara,
  Smallman, Sauer, Tarbutt, and Hinds}}]{hud11}
\bibinfo{author}{\bibfnamefont{J.}~\bibnamefont{Hudson}},
  \bibinfo{author}{\bibfnamefont{D.~M.} \bibnamefont{Kara}},
  \bibinfo{author}{\bibfnamefont{I.~J.} \bibnamefont{Smallman}},
  \bibinfo{author}{\bibfnamefont{B.~E.} \bibnamefont{Sauer}},
  \bibinfo{author}{\bibfnamefont{M.~R.} \bibnamefont{Tarbutt}},
  \bibnamefont{and} \bibinfo{author}{\bibfnamefont{E.~A.} \bibnamefont{Hinds}},
  \bibinfo{journal}{Nature} \textbf{\bibinfo{volume}{473}},
  \bibinfo{pages}{493} (\bibinfo{year}{2011}).

\bibitem[{\citenamefont{Jarzynski}(1997)}]{jar97}
\bibinfo{author}{\bibfnamefont{C.}~\bibnamefont{Jarzynski}},
  \bibinfo{journal}{\PRL} \textbf{\bibinfo{volume}{78}}, \bibinfo{pages}{2690}
  (\bibinfo{year}{1997}).

\bibitem[{\citenamefont{Kawai et~al.}(2007)\citenamefont{Kawai, Parrondo, and
  den Broeck}}]{kaw07}
\bibinfo{author}{\bibfnamefont{R.}~\bibnamefont{Kawai}},
  \bibinfo{author}{\bibfnamefont{J.~M.~R.} \bibnamefont{Parrondo}},
  \bibnamefont{and} \bibinfo{author}{\bibfnamefont{C.~V.} \bibnamefont{den
  Broeck}}, \bibinfo{journal}{\PRL} \textbf{\bibinfo{volume}{98}},
  \bibinfo{pages}{080602} (\bibinfo{year}{2007}).

\bibitem[{\citenamefont{Jarzynski}(2008)}]{jar08}
\bibinfo{author}{\bibfnamefont{C.}~\bibnamefont{Jarzynski}},
  \bibinfo{journal}{Eur. Phys. J. B} \textbf{\bibinfo{volume}{64}},
  \bibinfo{pages}{331} (\bibinfo{year}{2008}).

\bibitem[{\citenamefont{Esposito and van~den Broeck}(2011)}]{esp11}
\bibinfo{author}{\bibfnamefont{M.}~\bibnamefont{Esposito}} \bibnamefont{and}
  \bibinfo{author}{\bibfnamefont{C.}~\bibnamefont{van~den Broeck}},
  \bibinfo{journal}{\EPL} \textbf{\bibinfo{volume}{95}}, \bibinfo{pages}{40004}
  (\bibinfo{year}{2011}).

\bibitem[{\citenamefont{Talkner et~al.}(2007)\citenamefont{Talkner, Lutz, and
  H\"anggi}}]{tal07}
\bibinfo{author}{\bibfnamefont{P.}~\bibnamefont{Talkner}},
  \bibinfo{author}{\bibfnamefont{E.}~\bibnamefont{Lutz}}, \bibnamefont{and}
  \bibinfo{author}{\bibfnamefont{P.}~\bibnamefont{H\"anggi}},
  \bibinfo{journal}{\PRE} \textbf{\bibinfo{volume}{75}}, \bibinfo{pages}{050102
  (R)} (\bibinfo{year}{2007}).

\bibitem[{\citenamefont{Campisi
  et~al.}(2011{\natexlab{a}})\citenamefont{Campisi, H\"anggi, and
  Talkner}}]{cam11}
\bibinfo{author}{\bibfnamefont{M.}~\bibnamefont{Campisi}},
  \bibinfo{author}{\bibfnamefont{P.}~\bibnamefont{H\"anggi}}, \bibnamefont{and}
  \bibinfo{author}{\bibfnamefont{P.}~\bibnamefont{Talkner}},
  \bibinfo{journal}{\RMP} \textbf{\bibinfo{volume}{83}}, \bibinfo{pages}{771}
  (\bibinfo{year}{2011}{\natexlab{a}}).

\bibitem[{\citenamefont{Vedral}(2012)}]{ved12}
\bibinfo{author}{\bibfnamefont{V.}~\bibnamefont{Vedral}}, \bibinfo{journal}{J.
  Phys. A: Math. Theor.} \textbf{\bibinfo{volume}{45}}, \bibinfo{pages}{272001}
  (\bibinfo{year}{2012}).

\bibitem[{\citenamefont{Yuen and Ozawa}(1993)}]{yue93}
\bibinfo{author}{\bibfnamefont{H.~P.} \bibnamefont{Yuen}} \bibnamefont{and}
  \bibinfo{author}{\bibfnamefont{M.}~\bibnamefont{Ozawa}},
  \bibinfo{journal}{\PRL} \textbf{\bibinfo{volume}{70}}, \bibinfo{pages}{363}
  (\bibinfo{year}{1993}).

\bibitem[{\citenamefont{Hall and O'Rourke}(1993)}]{hal93}
\bibinfo{author}{\bibfnamefont{M.~J.~W.} \bibnamefont{Hall}} \bibnamefont{and}
  \bibinfo{author}{\bibfnamefont{M.~J.} \bibnamefont{O'Rourke}},
  \bibinfo{journal}{Quantum Opt.} \textbf{\bibinfo{volume}{5}},
  \bibinfo{pages}{161} (\bibinfo{year}{1993}).

\bibitem[{\citenamefont{Fuchs and Caves}(1994)}]{fuc94}
\bibinfo{author}{\bibfnamefont{C.~A.} \bibnamefont{Fuchs}} \bibnamefont{and}
  \bibinfo{author}{\bibfnamefont{C.~M.} \bibnamefont{Caves}},
  \bibinfo{journal}{\PRL} \textbf{\bibinfo{volume}{73}}, \bibinfo{pages}{3047}
  (\bibinfo{year}{1994}).

\bibitem[{\citenamefont{Hausladen et~al.}(1996)\citenamefont{Hausladen, Jozsa,
  Schumacher, Westmoreland, and Wootters}}]{hau96}
\bibinfo{author}{\bibfnamefont{P.}~\bibnamefont{Hausladen}},
  \bibinfo{author}{\bibfnamefont{R.}~\bibnamefont{Jozsa}},
  \bibinfo{author}{\bibfnamefont{B.}~\bibnamefont{Schumacher}},
  \bibinfo{author}{\bibfnamefont{M.}~\bibnamefont{Westmoreland}},
  \bibnamefont{and} \bibinfo{author}{\bibfnamefont{W.~K.}
  \bibnamefont{Wootters}}, \bibinfo{journal}{\PRA}
  \textbf{\bibinfo{volume}{54}}, \bibinfo{pages}{1869} (\bibinfo{year}{1996}).

\bibitem[{\citenamefont{Schumacher and Westmoreland}(1997)}]{sch97}
\bibinfo{author}{\bibfnamefont{B.}~\bibnamefont{Schumacher}} \bibnamefont{and}
  \bibinfo{author}{\bibfnamefont{M.~D.} \bibnamefont{Westmoreland}},
  \bibinfo{journal}{\PRA} \textbf{\bibinfo{volume}{56}}, \bibinfo{pages}{131}
  (\bibinfo{year}{1997}).

\bibitem[{\citenamefont{Schumacher et~al.}(1996)\citenamefont{Schumacher,
  Westmoreland, and Wootters}}]{sch962}
\bibinfo{author}{\bibfnamefont{B.}~\bibnamefont{Schumacher}},
  \bibinfo{author}{\bibfnamefont{M.}~\bibnamefont{Westmoreland}},
  \bibnamefont{and} \bibinfo{author}{\bibfnamefont{W.~K.}
  \bibnamefont{Wootters}}, \bibinfo{journal}{Phys. Rev. Lett.}
  \textbf{\bibinfo{volume}{76}}, \bibinfo{pages}{3452} (\bibinfo{year}{1996}).

\bibitem[{\citenamefont{Jacobs}(2003)}]{jac03}
\bibinfo{author}{\bibfnamefont{K.}~\bibnamefont{Jacobs}},
  \bibinfo{journal}{\PRA} \textbf{\bibinfo{volume}{68}},
  \bibinfo{pages}{054302} (\bibinfo{year}{2003}).

\bibitem[{\citenamefont{Jacobs}(2006)}]{jac06}
\bibinfo{author}{\bibfnamefont{K.}~\bibnamefont{Jacobs}}, \bibinfo{journal}{J.
  Math. Phys.} \textbf{\bibinfo{volume}{47}}, \bibinfo{pages}{012102}
  (\bibinfo{year}{2006}).

\bibitem[{\citenamefont{Schumacher}(1996)}]{sch96}
\bibinfo{author}{\bibfnamefont{B.}~\bibnamefont{Schumacher}},
  \bibinfo{journal}{\PRA} \textbf{\bibinfo{volume}{2614}}, \bibinfo{pages}{54}
  (\bibinfo{year}{1996}).

\bibitem[{\citenamefont{Talkner et~al.}(2009)\citenamefont{Talkner, Campisi,
  and H\"anggi}}]{tal09}
\bibinfo{author}{\bibfnamefont{P.}~\bibnamefont{Talkner}},
  \bibinfo{author}{\bibfnamefont{M.}~\bibnamefont{Campisi}}, \bibnamefont{and}
  \bibinfo{author}{\bibfnamefont{P.}~\bibnamefont{H\"anggi}},
  \bibinfo{journal}{J. Stat. Mech.} p. \bibinfo{pages}{P02025}
  (\bibinfo{year}{2009}).

\bibitem[{\citenamefont{Deffner and Lutz}(2011)}]{def11}
\bibinfo{author}{\bibfnamefont{S.}~\bibnamefont{Deffner}} \bibnamefont{and}
  \bibinfo{author}{\bibfnamefont{E.}~\bibnamefont{Lutz}},
  \bibinfo{journal}{\PRL} \textbf{\bibinfo{volume}{107}},
  \bibinfo{pages}{140404} (\bibinfo{year}{2011}).

\bibitem[{\citenamefont{Campisi
  et~al.}(2011{\natexlab{b}})\citenamefont{Campisi, Talkner, and
  H\"anggi}}]{cam11a}
\bibinfo{author}{\bibfnamefont{M.}~\bibnamefont{Campisi}},
  \bibinfo{author}{\bibfnamefont{P.}~\bibnamefont{Talkner}}, \bibnamefont{and}
  \bibinfo{author}{\bibfnamefont{P.}~\bibnamefont{H\"anggi}},
  \bibinfo{journal}{Phys. Rev. E} \textbf{\bibinfo{volume}{83}},
  \bibinfo{pages}{041114} (\bibinfo{year}{2011}{\natexlab{b}}).

\bibitem[{\citenamefont{Jarzynski and W\'ojcik}(2004)}]{woj04}
\bibinfo{author}{\bibfnamefont{C.}~\bibnamefont{Jarzynski}} \bibnamefont{and}
  \bibinfo{author}{\bibfnamefont{D.~K.} \bibnamefont{W\'ojcik}},
  \bibinfo{journal}{Phys. Rev. Lett.} \textbf{\bibinfo{volume}{92}},
  \bibinfo{pages}{230602} (\bibinfo{year}{2004}).

\bibitem[{\citenamefont{Morikuni and Tasaki}(2011)}]{mor11}
\bibinfo{author}{\bibfnamefont{Y.}~\bibnamefont{Morikuni}} \bibnamefont{and}
  \bibinfo{author}{\bibfnamefont{H.}~\bibnamefont{Tasaki}},
  \bibinfo{journal}{J. Stat. Phys.} \textbf{\bibinfo{volume}{143}},
  \bibinfo{pages}{1} (\bibinfo{year}{2011}).

\bibitem[{\citenamefont{Chandler}(1987)}]{chandler_87}
\bibinfo{author}{\bibfnamefont{D.}~\bibnamefont{Chandler}},
  \emph{\bibinfo{title}{Introduction to Modern Statistical Mechanics}}
  (\bibinfo{publisher}{Oxford University Press}, \bibinfo{address}{Oxford, UK},
  \bibinfo{year}{1987}).

\bibitem[{\citenamefont{Kurchan}(2000)}]{kur00}
\bibinfo{author}{\bibfnamefont{J.}~\bibnamefont{Kurchan}}
  (\bibinfo{year}{2000}), \bibinfo{note}{arXiv:cond-mat/0007360v2.}

\bibitem[{\citenamefont{Tasaki}(2000)}]{tas00}
\bibinfo{author}{\bibfnamefont{H.}~\bibnamefont{Tasaki}}
  (\bibinfo{year}{2000}), \bibinfo{note}{arXiv:cond-mat/0000244v2}.

\bibitem[{\citenamefont{Braginsky and Khalili}(1992)}]{bra92}
\bibinfo{author}{\bibfnamefont{V.~B.} \bibnamefont{Braginsky}}
  \bibnamefont{and} \bibinfo{author}{\bibfnamefont{F.~Y.}
  \bibnamefont{Khalili}}, \emph{\bibinfo{title}{Quantum Measurement}}
  (\bibinfo{publisher}{Cambridge University Press},
  \bibinfo{address}{Cambridge, UK}, \bibinfo{year}{1992}).

\bibitem[{\citenamefont{Nielsen and Chuang}(2000)}]{nielsen_00}
\bibinfo{author}{\bibfnamefont{M.~A.} \bibnamefont{Nielsen}} \bibnamefont{and}
  \bibinfo{author}{\bibfnamefont{I.~L.} \bibnamefont{Chuang}},
  \emph{\bibinfo{title}{Quantum Computation and Quantum Information}}
  (\bibinfo{publisher}{Cambridge University Press},
  \bibinfo{address}{Cambridge, UK}, \bibinfo{year}{2000}).

\bibitem[{\citenamefont{Kullback}(1978)}]{kullback_78}
\bibinfo{author}{\bibfnamefont{S.}~\bibnamefont{Kullback}},
  \emph{\bibinfo{title}{Information Theory and Statistics}}
  (\bibinfo{publisher}{Peter Smith}, \bibinfo{address}{Gloucester, USA},
  \bibinfo{year}{1978}).

\bibitem[{\citenamefont{Golden}(1965)}]{gol65}
\bibinfo{author}{\bibfnamefont{S.}~\bibnamefont{Golden}},
  \bibinfo{journal}{\PR} \textbf{\bibinfo{volume}{137}}, \bibinfo{pages}{B1127}
  (\bibinfo{year}{1965}).

\bibitem[{\citenamefont{Thompson}(1965)}]{tho65}
\bibinfo{author}{\bibfnamefont{C.~J.} \bibnamefont{Thompson}},
  \bibinfo{journal}{J. Math. Phys.} \textbf{\bibinfo{volume}{6}},
  \bibinfo{pages}{1812} (\bibinfo{year}{1965}).

\bibitem[{\citenamefont{Ruskai}(1972)}]{rus72}
\bibinfo{author}{\bibfnamefont{M.~B.} \bibnamefont{Ruskai}},
  \bibinfo{journal}{Comm. Math. Phys.} \textbf{\bibinfo{volume}{26}},
  \bibinfo{pages}{280} (\bibinfo{year}{1972}).

\bibitem[{\citenamefont{Ruskai}(2002)}]{rus02}
\bibinfo{author}{\bibfnamefont{M.~B.} \bibnamefont{Ruskai}},
  \bibinfo{journal}{J. Math. Phys.} \textbf{\bibinfo{volume}{9}},
  \bibinfo{pages}{4358} (\bibinfo{year}{2002}).

\bibitem[{\citenamefont{Petz}(1988)}]{pet88}
\bibinfo{author}{\bibfnamefont{D.}~\bibnamefont{Petz}},
  \bibinfo{journal}{Communications in Mathematical Physics}
  \textbf{\bibinfo{volume}{114}}, \bibinfo{pages}{345} (\bibinfo{year}{1988}).

\bibitem[{\citenamefont{Petz}(1994)}]{pet94}
\bibinfo{author}{\bibfnamefont{D.}~\bibnamefont{Petz}}, in
  \emph{\bibinfo{booktitle}{Functional Analysis and Operator Theory}}
  (\bibinfo{publisher}{Banach Center Publications}, \bibinfo{year}{1994}),
  vol.~\bibinfo{volume}{30}.

\bibitem[{\citenamefont{Sagawa and Ueda}(2010)}]{sag10}
\bibinfo{author}{\bibfnamefont{T.}~\bibnamefont{Sagawa}} \bibnamefont{and}
  \bibinfo{author}{\bibfnamefont{M.}~\bibnamefont{Ueda}},
  \bibinfo{journal}{\PRL} \textbf{\bibinfo{volume}{104}},
  \bibinfo{pages}{090602} (\bibinfo{year}{2010}).

\bibitem[{\citenamefont{Sagawa and Ueda}(2012)}]{sag12}
\bibinfo{author}{\bibfnamefont{T.}~\bibnamefont{Sagawa}} \bibnamefont{and}
  \bibinfo{author}{\bibfnamefont{M.}~\bibnamefont{Ueda}},
  \bibinfo{journal}{\PRE} \textbf{\bibinfo{volume}{85}},
  \bibinfo{pages}{021104} (\bibinfo{year}{2012}).

\end{thebibliography}

\end{document}